\documentclass[aps,nofootinbib,superscriptaddress, showpacs,preprintnumbers,  nofootinbibt,twocolumn]{revtex4-2}
\usepackage{eurosym}
\usepackage{amsmath}
\usepackage{bm}
\usepackage{amsfonts}
\usepackage{amssymb}
\usepackage{graphicx}

\def\be{\begin{equation}}
\def\ee{\end{equation}}
\def\bea{\begin{eqnarray}}
\def\eea{\end{eqnarray}}

\newcommand{\cH}{\ensuremath{\mathcal{H}}}

\begin{document}

\title{Conformal gravitational theories in the Barthel-Kropina type Finslerian geometry, and their cosmological implications}
\author{Rattanasak Hama}
\email{rattanasak.h@psu.ac.th}
\affiliation{Faculty of Science and Industrial Technology, Prince of Songkla University,
Surat Thani Campus, Surat Thani, 84000, Thailand,}
\author{Tiberiu Harko}
\email{tiberiu.harko@aira.astro.ro}
\affiliation{Department of Theoretical Physics, National Institute of Physics
and Nuclear Engineering (IFIN-HH), Bucharest, 077125 Romania,}
\affiliation{Department of Physics, Babes-Bolyai University, Kogalniceanu Street,
	Cluj-Napoca, 400084, Romania,}
\affiliation{Astronomical Observatory, 19 Ciresilor Street,
Cluj-Napoca 400487, Romania,}
\author{Sorin V. Sabau}
\email{sorin@tokai.ac.jp}
\affiliation{School of Biological Sciences, Department of Biology, Tokai University, Sapporo 005-8600, Japan,}
\affiliation{Graduate School of Science and Technology, Physical and Mathematical Sciences, \\
Tokai University, Sapporo 005-8600, Japan}

\begin{abstract}
We consider dark energy models obtained from the general conformal transformation of the Kropina metric, representing an $(\alpha, \beta)$ type Finslerian geometry, constructed as the ratio of the square of a Riemannian metric $\alpha$, and of the one-form $\beta$. Conformal symmetries do appear in many fields of physics, and they may play a fundamental role in the understanding of the Universe. We investigate the possibility of obtaining conformal theories of gravity in the osculating Barthel-Kropina geometric framework, where gravitation is described by an extended Finslerian type model, with the metric tensor depending on both the base space coordinates, and on a vector field. We show that it is possible to formulate a family of conformal Barthel-Kropina theories in an osculating geometry with second-order field equations, depending on the properties of the conformal factor, whose presence leads to the appearance of an effective scalar field, of geometric origin, in the gravitational field equations. The cosmological implications of the theory are investigated in detail, by assuming a specific relation between the component of the one-form of the Kropina metric, and the conformal factor. The cosmological evolution is thus determined by the initial conditions of the scalar field, and a free parameter of the model. We analyze in detail three cosmological models, corresponding to different values of the theory parameters. Our results show that the conformal Barthel-Kropina model could give an acceptable description of the observational data, and may represent a theoretically attractive alternative to the standard $\Lambda$CDM cosmology.
\end{abstract}

\date{\today }
\maketitle


\section{Introduction}

Einstein’s theory of General Relativity (GR) represents an impressive scientific achievement of the last century. Its exceptional success is mainly due
to its remarkable geometric description of gravity \cite{W1,W2}. GR is a far-reaching theory of spacetime, matter, and gravity, which gives a very precise account of the dynamics of the Solar System. However, when extended to very large, and very small scales, the theory faces several important problems, mainly coming out from cosmology
and quantum field theory. Moreover, several recent cosmological observations have raised serious concerns about the validity of GR as the theoretical foundation of cosmology.

The discovery of the recent acceleration of the Universe \cite{A1,A2,A3} can be explained very well by reinserting into the Einstein gravitational field equations the old, and much debated, cosmological constant $\Lambda$ \cite{Ein}, together with a mysterious (and not yet understood) matter type component, called dark matter. Dark matter is supposed to be pressureless, and cold. The $\Lambda$CDM ($\Lambda$ Cold Dark Matter) cosmological model has become in the recent years the standard paradigmatic approach for the interpretation of the cosmological data, and, in this respect, it is extremely successful.  However, there are a few important open questions that may suggest that $\Lambda$CDM is just a first approximation of a more realistic model, yet to be found \cite{R1}. Firstly, it lacks a firm theoretical basis (no generally accepted explanation of the geometrical or physical character of the cosmological constant is known), and, secondly, despite the extremely intensive experimental and observational effort, then particles assumed to form dark matter have not been discovered yet in terrestrial experiments, or astrophysical observations.

The tremendous increase of the precision of the recent cosmological observations, and of the technological advances in the field led to an other important challenge the $\Lambda$CDM standard  paradigm must face.  There are significant deviations between the Hubble expansion rates of the Universe as measured by the Planck satellite experiment by using the Cosmic Microwave Background Radiation (CMBR), originating from the early Universe as a result of the decoupling of matter and radiation, and the low redshift (local) measurements. These differences in the values of the present day Hubble constant $H_0$ are generally called Hubble tension, which could represent a fundamental  crisis in cosmology \cite{O2,O2a,O1,O3,O4,O5,O6}. The difference in the determination of the numerical values of H0, as obtained by the Planck satellite,  $H_0=66.93 \pm 0.62$ km/ s/ Mpc \cite{O5,O6}, and the values of $H_0=73.24\pm 1.74$ km/ s/ Mpc \cite{O1} determined by the SH0ES collaboration, exceeds 3$\sigma$ \cite{O6}. The Hubble tension, if it indeed exists, strongly points towards the necessity of finding new gravitational theories, as well as  of the essential requirement of the replacement of the $\Lambda$CDM model with an alternative one.

The Big Bang singularity, so important for the understanding of the nature of the Universe, is still unexplained in the framework of the $\Lambda$CDM cosmology, and it seems  that GR can not describe the Universe at extremely high density phases, and in the presence of very strong gravitational fields. On the other hand, there is very little progress, if any, in the quantization of spacetime, geometry, and gravity \cite{R2}. Since a quantum description of the gravitational interaction is (still) missing, and there is no complete quantum formulation of gravity, GR cannot be considered yet as a fundamental physical theory, similar to the other theories of physics describing so successfully elementary particle interactions.

The solution of these fundamental problems may require the consideration of novel theories of gravity, which contain GR as a particular weak field limit. There are a large number of attempts for constructing gravitational theories that are alternatives to GR, and they are constructed by using various, and sometimes very different, mathematical and physical perspectives (for comprehensive and detailed reviews of modified gravity theories, and their astrophysical and cosmological applications, see \cite{Rev0, Rev1,Rev2,Rev3}.

One of the interesting approaches to gravitational phenomena is related to the use of the conformal transformations (rescalings) \cite{CT1,CT2,CT3,CT4,CT5,CT6,CT7,CT8,CT9,CT10}. The important role the conformal transformations and structures may play in gravitational physics and  cosmology was suggested by Penrose \cite{CT1}, who developed an interesting cosmological model named Conformal Cyclic Cosmology (CCC). This theoretical model originated from the consideration of the fact that when the de Sitter exponentially accelerating stage, triggered by the existence of the positive $\Lambda$, ends, the spacetime  is conformally flat, and space-like. This geometry coincides with the initial boundary of the very early  Universe immediately after the Big Bang. In
the CCC model, one assumes that the Universe is made up of eons, which represent time oriented manifolds, with the conformally invariant compactifications possessing space-like null infinities. The CCC
model  has been studied in detail in \cite{CT2,CT3,CT4,CT5,CT6,CT7,CT8}.

The important role of the local conformal symmetry transformations  has been pointed out by ’t Hooft in \cite{CT9}. where it has been shown that conformal symmetry is an exact symmetry of nature, which is broken spontaneously. The breaking of the conformal symmetry could reveal a physical process explaining the small-scale properties of gravity. Conformal symmetry could be of equal importance as the Lorentz symmetry of the fundamental equations and laws of the elementary particle physics, and it may significantly contribute to the understanding of the physics of the Planck scale. By supposing that local conformal symmetry is an exact, but spontaneously broken symmetry of the physical world, a theory of the gravitational interaction was proposed
in \cite{CT10}. In this theory the conformal part of the metric is interpreted as a dilaton field. The theory has interesting physical consequences, with the black holes transformed into topologically trivial, regular solitons, without singularities, firewalls, or horizons.

By using conformal transformations of the metric, and of the physical and geometrical quantities,  it is possible to reduce gravitational theories containing higher-order and nonminimally coupled terms to GR plus some minimally coupled scalar fields \cite{C1, C2, C3}. Hence, in the framework of general relativity, it is achievable to change frames via conformal transformations. Two frames obtained from each other by conformal transformations are called conformal frames. It is important to point out that conformal frames are  mathematically equivalent \cite{C1,C2,C3}. However, their physical equivalence is a topic that generated strong debates among  physicists \cite{C2,C3}. Among the many possible conformal frames, two of them are of special interest, and they are called the Einstein frame, and the Jordan frame, respectively. In the Einstein frame there are only minimal coupling terms in the action. On the other hand, in the Jordan frame, non-minimal couplings between the gravitational fields, described by equivalent geometric quantities, and the other fields are present \cite{C3}.

The conformal transformations can also be interpreted, both mathematically and physically, as local unit transformations (rescaling of the lengths and distances). They were first considered by Hermann Weyl in his proposal for a unified theory of gravitation and electromagnetism  \cite{We1,We2,We3}, in which he also introduced the first generalization of the Riemann geometry. Weyl called the conformal transformations gauge transformations. Gauge transformations have become the standard theoretical tool in elementary particle physics, and gauge field theories are the basis of our present day understanding of the properties of the elementary particles. On the other hand, in the field of gravitational theories, the Weyl gauge transformations are called conformal transformations, and the invariance under them of physical laws, or geometric quantities, is called conformal invariance \cite{C1,C2, C3}.  The  electromagnetic field equations satisfy the local scale invariance.  At this moment one should note that local scale transformations do not keep the magnitudes of the vectors constant as they are parallelly displaced in the spacetime manifold. In Riemannian geometry, a nonvanishing curvature implies that the direction of a vector, parallelly transported around a closed path, is modified with respect to the direction of the initial vector, while its length remains unchanged. On the other hand, in Weyl geometry, the length of a vector is modified when parallelly transported around a closed loop, with the change being a function of the spacetime position. Gravitational models and theories based on the Weyl geometry have been extensively investigated and studied in the mathematical and physical literature \cite{Ma0,Ma4,Ma5,Q1, Q2, Q10, Q17,Q20, Q23, Q24, Laz,C0a,C1a,C2a,C3a,C4a, Gh1,Gh2, Gh6,Gh7,Gh8,Gh9,Gh10,Ca1,Ca2,Ca3,Ca4,Ca5,Ca6,Ca7, Bera,Berb,Berc,Berd}.

In the same year Weyl proposed his beautiful generalization of Riemann geometry, another important geometric theory was published. This is called Finsler geometry \cite{F1}, and it also represents an important extension of Riemann geometry. Even if Chern stated that Finsler geometry is  ”... just Riemannian geometry without the quadratic restriction”, in the following we will still refer to Finsler geometry as a generalization of Rimeann geometry. Actually, the geometry of Finsler was already predicted by Riemann \cite{r1}, who defined a geometric structure in a general space as given by the expression $ds=F\left(x^1,...,x^n; dx^1,...,dx^n\right)=F(x,dx)$. In this definition, for a nonzero $y$, $y\neq 0$, the function $F(x, y)$, called the Finsler metric function,  must be a positive function defined on the tangent bundle $TM$. Moreover, $F(x,y)$ must satisfy the important requirement of being homogeneous of degree one in $y$, thus satisfying the condition $F(x,\nu dx)=\nu F(x,dx)$, where $\nu$ is a positive constant. For $F^2=g_{ij}(x)dx^idx^j$,  we obtain the important limiting case of the Riemann geometry \cite{F2}.

The Finsler metric function $F$ can be expressed by using the canonical coordinates  $(x, y) = \left(x^I,y^I\right)$ of the tangent bundle $TM$, where by $y = y^I\left(\partial /\partial x^I\right)$ we have denoted the tangent vector at the point $x$ of the base manifold. Thus, the arc element in a general Finsler space takes the form $ds^2=g_{IJ}dx^Idx^J$. Finsler spaces posses a much general mathematical and geometrical structure, as compared to the Riemann spaces. For example,  in a Finsler space it is possible to define three kinds of curvature tensors $\left(R_{\nu \lambda \mu}^{\kappa}, S_{\nu \lambda \mu}^{\kappa},P_{\nu \lambda \mu}^{\kappa} \right)$, while the number of torsion tensors is five \cite{Bao}.

Even that the first physical applications of the Finsler geometry have been proposed after a relatively long time after its birth, physical and gravitational theories based on Finsler geometries have been intensively investigated as alternatives, or extensions of standard GR \cite{Rand, R1a, R2a,R3a,R4a,R5a,R6a,R7a,R8a, Tave1,Tave2,Tave3,Tave4, Voicu2, Hor1b, Hor2b, Hor3b,Hor4b,As0,As1,As2,As3,Miron, Ikeda, Rutz, Lixin, Voicu1}. One of the interesting possibilities for modelling gravitational phenomena is the dark gravity approach, which goes beyond the geometrical and mathematical formalism of the Riemann spaces. In this direction,  Finsler type theories and cosmological models are important and interesting alternatives to the standard $\Lambda$CDM model, since they can provide a geometric explanation, or replacement, of dark energy, and perhaps even of dark matter. A large number of studies have been devoted to the investigation of the possible applications of the Finsler geometry in gravitational physics and cosmology, with the main goal of understanding from a new geometric perspective the evolution and the dynamics of the cosmic components  \cite{Fc1,Fc2,Fc3,Fc4,Fc5,Fc6,Fc7,Fc8,Fc9,Fc10,Fc11,Fc12,Fc13,Fc14,Fc15,Fc16,Fc17,Fc18,Fc19,Fc19a, Fc20,Fc21,Fc22,Fc23, Fc24, Fc24a, Fc25,Fc26, Fc27, Fc28}.

In this respect, the cosmological applications of a special class of Finsler geomeries, called the Barthel-Kropina geometries, have been investigated in detail in \cite{Fc27} and \cite{Fc28}, respectively. The Kropina spaces are $(\alpha, \beta)$ type Finsler spaces, in which the Finsler metric function is defined by $F=\alpha ^2/\beta$, where $\alpha$ is a Riemannian metric, $\alpha (x,y)=\left[g_{IJ}dx^I dx^J\right]^{1/2}$, and $\beta ((x,y)=A_I(x)dy^I$ is an one-form. To simplify the mathematical approach, one can use the theory of the osculating Riemann spaces of Finsler geometries \cite{Na1,Na2}.
In the osculating space approach one associates  to a complicated Finsler geometric object a simpler mathematical one, like, for example, a Riemann
metric. Hence, with the help of the osculating approach, a simpler mathematical formalism can be obtained. In the case of the Kropina metric, one can choose the field $Y(x)$ as $Y(x)=A(x)$, and one can define the A-osculating Riemannian manifold $\left(M, \hat{g}_{IJ}(x,A(x)\right)$. One can associate to this structure the Barthel connection, which is nothing but the Levi-Civita connection of the Riemann metric $\hat{g}_{IJ}(x)=\hat{g}_{IJ}(x,A(x)$.

For a cosmological metric of the Friedmann-Lemaitre-Robertson-Walker type, the generalized Friedmann equations in the Barthel-Kropina geometry have been obtained in \cite{Fc27}. These equations lead to a dark energy model, which can explain the accelerating expansion of the Universe, and other observational cosmological features. The predictions of the Barthel-Kropina dark energy model were  compared  with observational data in \cite{Fc28}.  The model parameters were constrained by using 57 Hubble data points, and the Pantheon Supernovae Type Ia data sample. The  statistical analysis was performed with the use of Markov Chain Monte Carlo (MCMC) numerical simulations. An in depth comparison with the standard $\Lambda$CDM model was also considered, and the Akaike Information Criterion (AIC), and the Bayesian Information Criterion (BIC) were considered as selection tools for the two models. The statefinder diagnostics were also considered. The obtained results show that the Barthel-Kropina dark energy model can give an excellent explanation of the cosmological observational data, and hence it may represent an
interesting and viable alternative to the $\Lambda$CDM model.

It is the goal of the present work to introduce, and develop, another view on the Barthel-Kropina cosmology. Namely, we will consider the effects of a conformal transformation on the Finsler function $F(x,y)$, of the form $F(x,y)\rightarrow e^{\sigma (x)}F(x,y)$. From a physical point of view such a transformation represents a change from the Finslerian Einstein frame to a Finslerian Jordan-type frame. As a result of the conformal transformation, the Levi-Civita connection becomes a Weyl type connection, and a new scalar degree of freedom, associated to the conformal factor $\sigma (x)$ does appear in the mathematical structure of the Einstein gravitational field equations. In order to obtain a consistent description of the gravitational phenomena we adopt as Finslerian metric the Kropina type $(\alpha, \beta)$ metric, and the osculating geometrical approach, in which we assume $y=Y(x)$. Moreover, we introduce the Barthel connection, which is the Levi-Civita connection associated to the Riemannian metric $g(x,Y(x)$.  As a first step in our analysis we obtain the expression of the Einstein tensor in the conformal Barthel-Kropina geometry. The vector $Y$ is assumed to be the coefficient of the one form $\beta=A_Idx^I$, $Y=A$. We also investigate the conformal transformation properties of the matter action. After formulating the gravitational field equations in a general form, by assuming the usual proportionality of the Einstein tensor with the matter energy-momentum tensor, the cosmological applications of the conformal Finsler type geometry are considered. By adopting for the Riemannian metric the Friedmann-Lemaitre-Robertson-Walker form, the generalized Friedmann equations are obtained, which also contain a new scalar degree of freedom, coming from the conformal factor, assumed to be a function of the cosmological time only. By assuming a specific relation between the coefficient of the one form $\beta$ and the conformal factor, we obtain a consistent cosmological model, formulated in the redshift space. The model is dependent on a single free parameter. A de Sitter type solution of the conformal Barthel-Kropina field equations does also exist. Several solution of the generalized Friedmann equations, corresponding to different values of the model parameter are obtained by using numerical methods, and the redshift evolution of the Hubble function, deceleration parameter, matter density, and conformal factor are obtained. A comparison with the Hubble observational data, and with the $\Lambda$CDM model is also performed. As a result of these invetigations it is found that the conformal Barthedl-Kropina model can give an acceptable description of the observational data, and it may represent an attractive alternative to standard cosmology.

The present paper is organized as follows. The conformal transformations in Riemann geometry, the Barthel-Kropiona cosmological model, and the conformal properties of the matter energy-momentum tensor  are briefly reviewed in Section~\ref{sect1}. The conformal transformation properties of the cosmological Kropina metric are considered in Section~\ref{sect2}, where the expression of the Einstein tensor is also obtained. The generalized Friedmann equations of the conformal Barthel-Kropina cosmological model are written down in Section~\ref{sect3}, where a dark energy model is also presented, built upon a specific form of the scalar conformal factor. Specific cosmological models are investigated numerically in Section~\ref{sect4}, where a comparison of the models with a limited set of observational data, and with the $\Lambda$CDM model, is also performed. We discuss and conclude our results in Section~\ref{sect5}.

\section{From conformal transformation in Riemann geometry to the Barthel-Kropina cosmology}\label{sect1}

In the following we will summarize some of the basic mathematical results on the conformal transformations in Riemann geometry to be used in the sequel. We will also present the basics of the Barthel-Kropina cosmological theory, and write down the generalized Friedmann equations for this model.

\subsection{Conformal transformations and Riemann geometry}

Let $\nabla$ be a connection on $M$. On $M$ we introduce a symmetric metric $g$ with the components of the metric tensor denoted by $g_ij$ The global formula of Levi-Civita connection is given by
\begin{equation}\label{eq_Levi-Civita}
(\nabla_{X}g)(Y,Z)=X(g(Y,Z))-g(\nabla_XY,Z)-g(Y,\nabla_XZ).
\end{equation}

Locally, we have
\begin{equation}\label{eq_local_nabla}
\nabla_{\frac{\partial}{\partial x^k}}\frac{\partial}{\partial x^j}=\gamma^i_{jk}(x)\frac{\partial}{\partial x^i},
\end{equation}
where $\gamma^i_{jk}$ are called the {\it Christoffel symbols}, and they are defined according to
\be
\gamma^i_{jk}=\frac{1}{2}g^{is}\left(\frac{\partial g_{js}}{\partial x^k}+\frac{\partial g_{ks}}{\partial x^j}-\frac{\partial g_{jk}}{\partial x^s}\right).
\ee

The global expression of the curvature operator is given by
\be
R(X,Y)Z=\nabla_X\nabla_YZ-\nabla_Y\nabla_XZ-\nabla_{[X,Y]}Z.
\ee

Locally, we have
\be
R^{\ i}_{j\ kl}=\frac{\partial \gamma^i_{jl}}{\partial x^k}-\frac{\partial \gamma^i_{jk}}{\partial x^l}+\gamma^s_{jl}\gamma^i_{sk}-\gamma^s_{jk}\gamma^i_{sl}.
\ee

We consider now the conformal transformation of the metric given by
\be
\tilde{g}_{ij}(x)=\Omega ^2(x)g_{ij}(x)=e^{2\sigma(x)}g_{ij}(x),
\ee
where $\sigma (x)$ is an arbitrary function of the coordinates $x$ defined on the space-time manifold $M$, and we have denoted the conformally transformed metric by $\tilde{g}_{ij}$. Then, we obtain the first result in the conformal Riemann geometry in the form of

{\bf Lemma 1.}
The relation of Riemannian Christoffel of $\tilde{g}_{ij}$ and $g_{ij}$ is
\begin{equation}\label{Lema1}
\begin{split}
\tilde{\gamma}^i_{jk}&=\gamma^i_{jk}+\delta^i_j\sigma_k+\delta^i_k\sigma_j-\sigma^i g_{jk},\\
\end{split}
\end{equation}
where
\be
\sigma_j:=\frac{\partial\sigma(x)}{\partial x^j}, \sigma^i=g^{ij}\sigma_j,
\ee

Next, we obtain for the definition of the covariant derivative the expression
\be
\tilde {\nabla }_{X}Y=\nabla _{X}Y+d\sigma (X)Y+d\sigma (Y)X-g(X,Y)\nabla \sigma,
\ee
where we have denoted
\be
\nabla\sigma=\sigma^i\frac{\partial}{\partial x^i}, d\sigma(x)=\frac{\partial\sigma(x)}{\partial x^i}X^i,
\ee
for any $X=X^i\frac{\partial}{\partial x^i}$. In local coordinates the proof of this relation is as follows
\begin{equation*}
\begin{split}
\tilde{\gamma}^i_{jk}\frac{\partial}{\partial x^i}
&=\gamma^i_{jk}\frac{\partial}{\partial x^i}+\frac{\partial\sigma}{\partial x^k}\frac{\partial}{\partial x^j}+\frac{\partial\sigma}{\partial x^j}\frac{\partial}{\partial x^k}-\sigma^i g_{jk}\frac{\partial}{\partial x^i}\\
\tilde{\nabla}_{\frac{\partial}{\partial x^k}}\frac{\partial}{\partial x^j}
&=\nabla_{\frac{\partial}{\partial x^k}}\frac{\partial}{\partial x^j}
+d\sigma\left(\frac{\partial}{\partial x^k}\right)\frac{\partial}{\partial x^j}
+d\sigma\left(\frac{\partial}{\partial x^j}\right)\frac{\partial}{\partial x^k}\\
&-g\left(\frac{\partial}{\partial x^j},\frac{\partial}{\partial x^k}\right)\sigma^i\frac{\partial}{\partial x^i}.
\end{split}
\end{equation*}

If we denote $X=\frac{\partial}{\partial x^k}$ and $Y=\frac{\partial}{\partial x^j}$, then
\begin{equation*}
\begin{split}
\tilde{\nabla}_XY
&=\nabla_XY
+d\sigma\left(X\right)Y
+d\sigma\left(Y\right)X
-g\left(Y,X\right)\sigma^i\frac{\partial}{\partial x^i}\\
&=\nabla_XY
+d\sigma\left(X\right)Y
+d\sigma\left(Y\right)X
-g\left(X,Y\right)\nabla\sigma.
\end{split}
\end{equation*}

Next, we consider the curvature properties of the conformally transformed metric.

{\bf Lemma 2.}
If we consider the conformal transformation $\tilde{g}_{ij}=e^{2\sigma(x)}g_{ij}(x)$ then the relation of Riemannian curvature of $\tilde{g}_{ij}$ and $g_{ij}$ is:
\begin{equation*}
\begin{split}
\tilde{R}_{ijkl}&=e^{2\sigma}R_{ijkl}-e^{2\sigma}\left(g_{ik}T_{jl}+g_{jl}T_{ik}-g_{il}T_{jk}-g_{jk}T_{il}\right),\\
\tilde{R}^{\ i}_{j\ kl}&=\tilde{g}^{is}\tilde{R}_{sjkl}
=R^{\ i}_{j\ kl}-\left(\delta^i_kT_{jl}+g_{jl}T^i_k-\delta^i_lT_{jk}-g_{jk}T^i_l\right),
\end{split}
\end{equation*}
where
\begin{equation*}
\begin{split}
\nabla_i \sigma&=\frac{\partial \sigma}{\partial x^i}=\sigma_i,
\nabla_i\nabla_j \sigma=\frac{\partial\sigma_i}{\partial x^j}-\sigma_p\gamma^p_{ij}=\sigma_{ij}-\sigma_p\gamma^p_{ij},\\
T_{ij}&=\nabla_i\nabla_j \sigma-\nabla_i \sigma\nabla_j \sigma+\frac{1}{2}|d\sigma|^2g_{ij}\\
&=\sigma_{ij}-\sigma_p\gamma^p_{ij}-\sigma_i\sigma_j+\frac{1}{2}|d\sigma|^2g_{ij}
,\\
T^i_j&=g^{is}T_{sj}=g^{is}\left(\sigma_{sj}-\sigma_p\gamma^p_{sj}-\sigma_s\sigma_j+\frac{1}{2}|d\sigma|^2g_{sj}\right),\\
\end{split}
\end{equation*}
and
\begin{equation}
|d\sigma|^2=\sigma^i\sigma_i, \sigma_{ij}=\frac{\partial^2\sigma}{\partial x^i\partial x^j}, \sigma^i_j=g^{is}\sigma_{sj}.
\end{equation}
Here $\nabla_i$ or $|i$ is the covariant derivative with respect to the Levi-Civita connection of $g$.


{\bf Remark.}
Please pay attention to the notation above, We have defined
$$
\sigma_{ij}:=\frac{\partial^2\sigma}{\partial x^i\partial x^j}\ \text{and}\ \sigma_j^k:=g^{ik}\sigma_{ij},
$$
which is different from
\begin{equation*}
\begin{split}
\frac{\partial\sigma^k}{\partial x^j}&=\frac{\partial}{\partial x^j}\left(g^{ik}\sigma_i\right)
=\frac{\partial g^{ik}}{\partial x^j}\sigma_i+g^{ik}\frac{\partial\sigma_i}{\partial x^j}\\
&=\frac{\partial g^{ik}}{\partial x^j}\sigma_i+g^{ik}\frac{\partial\sigma}{\partial x^i\partial x^j}
=\frac{\partial g^{ik}}{\partial x^j}\sigma_i+g^{ik}\sigma_{ij}.
\end{split}
\end{equation*}

{\bf Lemma 3.}
If we consider the conformal transformation $\tilde{g}_{ij}=e^{2\sigma(x)}g_{ij}(x)$ then the relation of Ricci tensor of $\tilde{g}_{ij}$ and $g_{ij}$ is:
\begin{equation*}
\begin{split}
\tilde{R}_{ij}=
&R_{ij}+(n-2)(\sigma_{ij}-\sigma_i\sigma_j-\sigma_m\gamma^m_{ij})\\
&+(\Delta\sigma+(n-2)|d\sigma|^2)g_{ij},
\end{split}
\end{equation*}
where
$$
\Delta \sigma=g^{jk}\frac{\partial^2\sigma}{\partial x^j\partial x^k}-g^{jk}\gamma^l_{jk}\frac{\partial \sigma}{\partial x^l}=g^{jk}\left(\sigma_{jk}-\gamma^l_{jk}\sigma_l\right).
$$

{\bf Lemma 4.}
If we consider the conformal transformation $\tilde{g}_{ij}=e^{2\sigma(x)}g_{ij}(x)$ then the relation between the scalar curvatures of $\tilde{g}_{ij}$ and $g_{ij}$ is:
\begin{equation*}
\begin{split}
\tilde{R}&=
e^{-2\sigma}\left[R+2(n-1)\Delta\sigma+(n-2)(n-1)|d\sigma|^2\right].
\end{split}
\end{equation*}

{\bf Lemma 5.}
If we consider the conformal transformation $\tilde{g}_{ij}=e^{2\sigma(x)}g_{ij}(x)$ then the Einstein tensor of $\tilde{g}_{ij}$ is:
\bea
\tilde{G}_{ij}&=&G_{ij}+(n-2)(\sigma_{ij}-\sigma_m\gamma^m_{ij}-\sigma_i\sigma_j) \nonumber\\
&&-(n-2)\big\{\Delta\sigma+\frac{(n-3)}{2}|d\sigma|^2\big\}g_{ij},
\eea
where
\begin{equation*}
\begin{split}
\tilde{G}_{ij}:=\tilde{R}_{ij}-\frac{1}{2}\tilde{R}\tilde{g}_{ij},
G_{ij}:=R_{ij}-\frac{1}{2}Rg_{ij}.
\end{split}
\end{equation*}

\subsubsection{Conformal transformation of matter}

Until now, in the previous Section, we have considered  only the conformal transformation properties of the geometrical
quantities. We turn now to the matter part of gravitational action. The matter action can be generally written as
\be
S_m=\int{L_m(g,\psi)\sqrt{-g}d^4x},
\ee
where the matter Lagrangian $L_m$ is assumed to be a function of the metric tensor and of the (bosonic or fermionic) matter fields $\psi$. To investigate the conformal properties of the matter Lagrangian density we assume first that under conformal transformations the matter Lagrangian is transformed according to the rule
\be\label{CL}
\tilde{L}_m=e ^{-4\sigma (x)}L_m.
\ee
Then the conformally tarnsformed action becomes
\bea
\tilde{S}_m&=&\int{\tilde{L}_m\sqrt{-\tilde{g}}d^4x}=\int{e^{-4\sigma (x)}L_m e^{4\sigma (x)}\sqrt{-g}d^4x}\nonumber\\
&=&S_m,
\eea

Hence, it follows that the action of the ordinary baryonic matter is invariant under the considered conformal transformations (\ref{CL}). This result implies that the baryonic matter component of the gravitational interaction can be described in all conformally related frames by the same expression, since it is an invariant quantity.

The matter energy-momentum tensor is defined according to the expression
\be
T_{I J}=\frac{2}{\sqrt{-g}}\frac{\delta }{\delta g^{I J}}\left(\sqrt{-g}L_m\right).
\ee

After performing a conformal transformation of the metric we obtain
\be
\tilde{T}_{I J}=e ^{-2 \sigma (x)} T_{I J}.
\ee
For the trace $\tilde{T}=\tilde{T}_I^I$ of the ordinary matter energy-momentum tensor we obtain the expression $\tilde{T}=e ^{-4\sigma (x)}T$, where $T=T_I^I$.

\subsection{The Barthel-Kropina cosmological model}\label{sec_Review_Kropina}

In \cite{Fc27} and \cite{Fc28} we have considered the Kropina metric \cite{YS2014}
$$
F=\frac{\alpha^2}{\beta}=\frac{g_{IJ}(x)y^Iy^J}{A_I(x)y^I},\ I,J=\{0,1,2,3\}
$$
with fundamental tensor
\bea\label{eq_Kropina Hessian}
\hat{g}_{IJ}(x,y)&=&\frac{2\alpha^2}{\beta^2}g_{IJ}(x)+\frac{3\alpha^4}{\beta^4}A_IA_J-\frac{4\alpha^2}{\beta^3}(y_IA_J+y_JA_I)\nonumber\\
&&+\frac{4}{\beta^2}y_Iy_J,
\eea
where $y_I:=g_{IJ}y^J$.

Let us consider for the Riemannian metric the expression
$$
(g_{IJ}(x))=\begin{pmatrix}
1 & 0 & 0 & 0\\
0 & -a^2(x^0) & 0 & 0\\
0 & 0 & -a^2(x^0) & 0\\
0 & 0 & 0 & -a^2(x^0)
\end{pmatrix},
$$
representing the homogeneous and isotropic, flat FLRW model, and $\beta=A_I(x)y^I=A_0(x)y^0$,
where
$$
(A_I(x))=(A_0,0,0,0)=(a(x^0)\eta(x^0),0,0,0)
$$
is a covariant vector field on $M$.

Moreover, we consider the preferred direction
$$
Y=Y^I\frac{\partial}{\partial x^I}=A^I\frac{\partial}{\partial x^I},
$$
where $ A^I:=g^{IJ}A_J$. In the case of the FLRW-metric it follows that
$$
(Y^I)\equiv(A^I)=(Y_I)=(A_I)=(a(x^0)\eta(x^0),0,0,0).
$$

We evaluate
\begin{equation*}
\begin{split}
\beta\vert_{y=A}&=[a(x^0)\eta(x^0)]^2\\
(h_{IJ})\vert_{y=A}&=\begin{pmatrix}
0 & 0 & 0 & 0\\
0 & -a^2(x^0) & 0 & 0\\
0 & 0 & -a^2(x^0) & 0\\
0 & 0 & 0 & -a^2(x^0)
\end{pmatrix},
\end{split}
\end{equation*}
where $h_{IJ}:=g_{IJ}(x)-\frac{y_I}{\alpha}\frac{y_J}{\alpha}$, $y_I:=g_{IJ}(x)y^J$.

By substitution in \eqref{eq_Kropina Hessian} we obtain the osculating Riemannian metric
\begin{equation}\label{eq_osculating_Kropina Hessian}
\begin{split}
&\hat{g}_{IJ}(x)=\hat{g}_{IJ}(x,y=A)
=\begin{pmatrix}
\frac{1}{a^2\eta^2} & 0 & 0 & 0\\
0 & -\frac{2}{\eta^2} & 0 & 0\\
0 & 0 & -\frac{2}{\eta^2} & 0\\
0 & 0 & 0 & -\frac{2}{\eta^2}
\end{pmatrix}
\end{split}
\end{equation}
and the non vanishing components of the Christoffel symbols of the second kind of \eqref{eq_osculating_Kropina Hessian} are
\begin{equation}\label{eq_Christoffel_Kropina}
\hat{\gamma}^I_{JK}=\begin{cases}
\hat{\gamma}^0_{00}&=-\dfrac{\eta\cH+\eta'}{\eta}, \vspace{0.5cm}\\
\hat{\gamma}^0_{ij}&=-\dfrac{2a^2\eta'}{\eta}\delta_{ij}, \vspace{0.5cm}\\
\hat{\gamma}^i_{0j}&=-\dfrac{\eta'}{\eta}\delta^i_j,
\end{cases}
\end{equation}
where $\cH=\frac{a'}{a}$.


Recall the formula of Ricci tensor from \cite{Fc27}
$$
\hat{R}_{IJ}=\begin{cases}
\hat{R}_{00}=\dfrac{3}{\eta^2}\left[\eta\eta''+\eta\eta'\cH-(\eta')^2\right],\vspace{0.5cm}\\
\hat{R}_{ij}=\dfrac{2a^2}{\eta^2}\left[3(\eta')^2-\eta\eta''-\eta\eta'\cH\right]\delta_{ij},\\
\end{cases}
$$
and Ricci scalar
$$
\hat{R}=6a^2\left(\eta\eta''+\eta\eta'\cH-2(\eta')^2\right).
$$

The Einstein field equations, given by $\hat{G}_{00}=\left(8\pi G/c^4\right)\hat{g}_{00}\rho c^2$, and $\hat{G}_{ii}=-\left(8\pi G/c^4\right)\hat{g}_{ii}p$, respectively, where $\rho$ and $p$ denote the matter energy density, and pressure, respectively, give the system of the generalized Friedmann equations
\be\label{Fr1}
\frac{3(\eta')^2}{\eta^2}=\frac{8 \pi G}{c^4}\frac{1}{a^2\eta ^2}\rho c^2,
\ee
and
\be\label{Fr2}
a^2\left[-3(\eta')^2+2\eta\eta''+2\cH\eta\eta'\right]=\frac{8\pi G}{c^4}p,
\ee
respectively. By substituting the term $-3\left(\eta '\right)^2$ by using Eq.~(\ref{Fr1}), Eq.~(\ref{Fr2}) takes the simple form
\be\label{Fr3}
2a\eta \frac{d}{dx^0}\left(\eta 'a\right)=\frac{8\pi G}{c^4}\left(\rho c^2+p\right).
\ee

The cosmological implications of this model have been investigated in detail in \cite{Fc27} and \cite{Fc28}, respectively.

\section{Conformal transformation and the Kropina metric}\label{sect2}

In the present Section we will consider the conformal transformation properties of the general $\alpha, \beta)$ metrics, with a special emphasis on the Kropina case. From a mathematical point of view the role of the conformal transformations in Finsler geometry, including the case of the $(\alpha, \beta)$ metrics, was investigated in \cite{Conf0, Conf1, Conf2,Conf3, Conf4,Conf5,Conf6}. From a physical point of view the role of the gauge transformations in Finsler geometry was considered in \cite{Conf7}.

\subsection{Conformal transformation of an $(\alpha, \beta)$ metric}

For any $(\alpha,\beta)$-metric $F=F(\alpha,\beta)$, we can consider its conformal transformation
\be\label{Cftrans}
\tilde{F}(x,y):=e^{\sigma(x)}F(x,y)=\tilde{F}(\tilde{\alpha},\tilde{\beta}),
\ee
which is an $(\tilde{\alpha},\tilde{\beta})$-metric, where
\be\label{Cftrans1}
\tilde{\alpha}=e^{\sigma(x)}\alpha,\ \tilde{\beta}=e^{\sigma(x)}\beta.
\ee

The fundamental tensor of $\tilde{F}$ is given by the Hessian
\be
\tilde{g}_{IJ}:=\frac{1}{2}\frac{\partial^2\tilde{F}^2}{\partial y^I\partial y^J}.
\ee
In the case of the general $(\alpha,\beta)$-metric we obtain
\bea
\hat{g}_{IJ}(x,y)&=&\rho g_{IJ}(x)+\rho_0b_Ib_J+\rho_1\left(b_I\frac{y_J}{\alpha}+b_J\frac{y_I}{\alpha}\right)\nonumber\\
&&-s\rho_1\frac{y_I}{\alpha}\frac{y_J}{\alpha},
\eea
where $y_I=g_{IJ}y^J$, i.e. $\frac{y_I}{\alpha}=\frac{\partial \alpha}{\partial y^I}$, and
\be
\rho=\phi^2-s\phi\phi', \rho_0=\phi\phi''+\phi'^2,\ \rho_1=-s(\phi\phi''+\phi'^2)+\phi\phi',
\ee
as function of $s=\frac{\beta}{\alpha}$. Here $I,J=\{0,1,2,3\}$.

By considering now a conformal transformation of $\tilde{F}$, we get
\bea\label{eq_conformal_Hessian}
\hat{\tilde{g}}_{IJ}(x,y)&=&\tilde{\rho} \tilde{g}_{IJ}(x)+\tilde{\rho}_0\tilde{b}_I\tilde{b}_J+\tilde{\rho}_1\left(\tilde{b}_I\frac{\tilde{y}_J}{\tilde{\alpha}}+\tilde{b}_J\frac{\tilde{y}_I}{\tilde{\alpha}}\right)\nonumber\\
&&-\tilde{s}\tilde{\rho}_1\frac{\tilde{y}_I}{\tilde{\alpha}}\frac{\tilde{y}_J}{\tilde{\alpha}}
=\rho e^{2\sigma} g_{IJ}(x)+\rho_0e^{2\sigma}b_Ib_J\nonumber\\
&&+\rho_1e^{2\sigma}\left(b_I\frac{y_J}{\alpha}+b_J\frac{y_I}{\alpha}\right)-s\rho_1e^{2\sigma}\frac{y_I}{\alpha}\frac{y_J}{\alpha}\nonumber\\
&=&e^{2\sigma} \hat{g}_{IJ}(x,y),
\eea
where we have used the relations
$$
\tilde{s}=\frac{\tilde{\beta}}{\tilde{\alpha}}=\frac{\beta}{\alpha},\ \tilde{\rho}=\rho,\ \tilde{\rho}_0=\rho_0,\ \tilde{\rho}_1=\rho_1,
$$
since they are just derivatives with respect to $s$.

Moreover,
$$
\frac{\tilde{y}_I}{\tilde{\alpha}}=\frac{\tilde{a}_{IJ}y^J}{\tilde{\alpha}}=\frac{e^{2\sigma}a_{IJ}y^J}{e^\sigma \alpha}=e^\sigma\frac{y_I}{\alpha}.
$$

\subsection{The Kropina case}\label{sec_Conformal_Kropina}

Likewise, we can extended the case  above by taking the conformal transform of the Kropina metric
$$
\tilde{F}:=e^{\sigma(x)}\frac{\alpha^2}{\beta}=\frac{\tilde{\alpha}^2}{\tilde{\beta}},
$$
where $\tilde{\alpha}=e^{\sigma(x)}\alpha$, $\tilde{\beta}=e^{\sigma(x)}\beta$.

Similarly with the standard Kropina case, the osculating Riemannian metric is obtained as
\be\label{14}
\hat{\tilde{g}}_{IJ}(x)=e^{2\sigma(x)}\hat{g}_{IJ}(x),
\ee
where $\hat{g}_{IJ}(x)$ is given by Eq.~\eqref{eq_osculating_Kropina Hessian}. Explicitly, the metric $\hat{g}_{IJ}(x)$ has the expression
\begin{equation}\label{conf_metr}
\begin{split}
&\hat{\tilde{g}}_{IJ}(x)=e^{2\sigma (x)}
\begin{pmatrix}
\frac{1}{a^2\eta^2} & 0 & 0 & 0\\
0 &- \frac{2}{\eta^2} & 0 & 0\\
0 & 0 & -\frac{2}{\eta^2} & 0\\
0 & 0 & 0 & -\frac{2}{\eta^2}
\end{pmatrix}
.
\end{split}
\end{equation}
\\
Now we can compute the Christoffel symbols, curvatures and the Einstein tensor for this conformal Riemannian metric.

\subsubsection{The generalized Einstein tensor in the conformal Barthel-Kropina model}

From Lemma 1, and by using Eq.~\eqref{eq_Christoffel_Kropina}, we obtain the components of the Christoffel symbols of the second kind of the conformal metric (\ref{14}) as
\begin{widetext}
\begin{equation}\label{eq_Christoffel_conformal_Kropina}
\hat{\tilde{\gamma}}^I_{JK}=\begin{cases}
\hat{\tilde{\gamma}}^0_{00}&=-\dfrac{\eta\cH+\eta'}{\eta}+\delta^0_0\sigma_0+\delta^0_0\sigma_0-\sigma^0\hat{g}_{00}= -\dfrac{\eta\cH+\eta'}{\eta}+\sigma_0,
\vspace{0.25cm}\\
\hat{\tilde{\gamma}}^0_{ij}&=-\dfrac{2a^2\eta'}{\eta}\delta_{ij}+\delta^0_{i}\sigma_j+\delta^0_j\sigma_i-\sigma^0\hat{g}_{ij}
=-2a^2\left(\dfrac{\eta'}{\eta}-\sigma_0\right)\delta_{ij},
 \vspace{0.25cm}\\
\hat{\tilde{\gamma}}^i_{0j}&=-\dfrac{\eta'}{\eta}\delta^i_j+\delta^i_0\sigma_j+\delta^i_j\sigma_0-\sigma^i\hat{g}_{0j}=\left(-\dfrac{\eta'}{\eta}+\sigma_0\right)\delta^i_j,
\vspace{0.25cm}\\
\hat{\tilde{\gamma}}^i_{jk}&=\delta^i_j\sigma_k+\delta^i_k\sigma_j-\delta^{im}\sigma_m\delta_{jk},
\vspace{0.25cm}\\
\hat{\tilde{\gamma}}^0_{0i}&=\sigma_i,
\vspace{0.25cm}\\
\hat{\tilde{\gamma}}^i_{00}&=-\dfrac{1}{2a^2}\delta^{im}\sigma_m,
\end{cases}
\end{equation}
\end{widetext}
where $\cH=\frac{a'}{a}$  and $\sigma^0=\hat{g}^{00}\sigma_0=a^2\eta^2\sigma_0$.

We now successively obtain
\begin{equation}
\begin{split}
|d\sigma|^2&=\sigma^I\sigma_I
=\hat{g}^{IJ}\sigma_J\sigma_I
=a^2\eta^2\sigma_0^2-\frac{\eta^2}{2}\sum_{i=1}^3\sigma_i^2,
\end{split}
\end{equation}
and
\begin{equation}
\begin{split}
\Delta\sigma=
a^2\eta^2\sigma_{00}
-\frac{\eta^2}{2}\sum_{i=1}^3\sigma_{ii}
-a^2\eta\left(2\eta'-\eta\cH\right)\sigma_0,
\end{split}
\end{equation}
respectively. Let us consider now the four-dimensional case with $n=4$. From Lemma 3, we obtain
$$
\hat{\tilde{R}}_{IJ}=
\hat{R}_{IJ}+2(\sigma_{IJ}-\sigma_I\sigma_J-\sigma_M\hat{\gamma}^M_{IJ})+(\Delta\sigma+2|d\sigma|^2)\hat{g}_{IJ},
$$
where $I,J,M=0,1,2,3$. We can formulate now the following

{\bf Lemma 6.}
\begin{enumerate}
\item $\sigma_M\hat{\gamma}^M_{00}=\sigma_0\hat{\gamma}^0_{00}+\sigma_i\hat{\gamma}^i_{00}=-\frac{\eta\cH+\eta'}{\eta}\sigma_0$.
\item $\sigma_M\hat{\gamma}^M_{ij}=\sigma_0\hat{\gamma}^0_{ij}+\sigma_t\hat{\gamma}^t_{ij}=-\frac{2a^2\eta'}{\eta}\sigma_0\delta_{ij}$.
\end{enumerate}
where $i,j=1,2,3$.

We proceed now to the computation of the components of the Ricci tensor. We obtain first
\bea
\hat{\tilde{R}}_{00}&=&3\left[\frac{\eta''+\eta'\cH}{\eta}-\frac{(\eta')^2}{\eta^2}
+\sigma_{00}+\cH\sigma_0\right]\nonumber\\
&&-\frac{1}{2a^2}\left(\sum_{k=1}^3\sigma_{kk}+2\sum_{k=1}^3\sigma_k^2\right),
\eea
and
\begin{equation}
\begin{split}
\hat{\tilde{R}}_{ij}=2\left(\sigma_{ij}-\sigma_i\sigma_j\right)+\psi_1\delta_{ij},
\end{split}
\end{equation}
respectively, where
\bea
\psi_1:&=&2a^2\Bigg\{\frac{1}{\eta^2}[3(\eta')^2-\eta\eta''-\eta\eta'\cH]
+\frac{1}{\eta}(4\eta'-\eta\cH)\sigma_0\nonumber\\
&&-\sigma_{00}-2\sigma_0^2\Bigg\}
+\sum_{k=1}^3\sigma_{kk}+2\sum_{k=1}^3\sigma_k^2.
\eea

On the other hand, if we consider $n=4$, from Lemma 4, we get
\be
\hat{\tilde{R}}=
e^{-2\sigma}\left[\hat{R}+6\Delta\sigma+6|d\sigma|^2\right].
\ee

Hence, we immediately obtain for $\hat{\tilde{R}}$ the expression
\bea
\hat{\tilde{R}}&=&6e^{-2\sigma}\Bigg\{a^2\left(\eta\eta''+\eta\eta'\cH-2(\eta')^2\right)\nonumber\\
&&+a^2\eta^2(\sigma_{00}+(\sigma_0)^2)
-\frac{\eta^2}{2}\left(\sum_{k=1}^2\sigma_{kk}+\sum_{k=1}^3\sigma_k^2\right)\nonumber\\
&&-a^2\eta\left(2\eta'-\eta\cH\right)\sigma_0
\Bigg\}.
\eea

Next, we consider the expression of the Einstein tensor of $\tilde{g}_{IJ}(x)=e^{2\sigma(x)}\hat{g}_{IJ}(x)$ in the four-dimensional case, $I,J=0,1,2,3$.

From Lemma 5, we obtain the general expression of the Einstein tensor
\bea
\hat{\tilde{R}}_{IJ}-\frac{1}{2}\hat{\tilde{R}}\hat{\tilde{g}}_{IJ}
&=&\hat{R}_{IJ}-\frac{1}{2}\hat{R}\hat{g}_{IJ}
+2(\sigma_{IJ}-\sigma_I\sigma_J-\sigma_M\hat{\gamma}^M_{IJ})\nonumber\\
&&-2(\Delta\sigma+\frac{1}{2}|d\sigma|^2)\hat{g}_{IJ}.
\eea

Then in the case of $(I,J)=(0,0)$, we get
\bea
\hat{\tilde{R}}_{00}-\frac{1}{2}\hat{\tilde{R}}\tilde{g}_{00}
&=&\hat{R}_{00}-\frac{1}{2}\hat{R}\hat{g}_{00}
+2\left[\sigma_{00}-(\sigma_0)^2-\sigma_M\hat{\gamma}^M_{00}\right]\nonumber\\
&&-2\left(\Delta\sigma+\frac{1}{2}|d\sigma|^2\right)\hat{g}_{00}.
\eea

Explicitly, for $\hat{\tilde{G}}_{00}$ we obtain
\begin{equation}
\begin{split}
\hat{\tilde{G}}_{00}&=\frac{3(\eta')^2}{\eta^2}
-3\sigma_0^2
+\frac{6\eta'}{\eta}\sigma_0
+\frac{1}{a^2}\left(\sum_{i=1}^3\sigma_{ii}+\frac{1}{2}\sum_{i=1}^3\sigma_i^2\right).
\end{split}
\end{equation}

For the spatial components of the Einstein tensor we find the expression
\begin{equation}\label{41}
\hat{\tilde{G}}_{ij}
=2(\sigma_{ij}-\sigma_i\sigma_j)+\psi_2\delta_{ij},
\end{equation}
where
\bea
\psi_2:&=&\frac{2a^2}{\eta^2}[-3(\eta')^2+2\eta\eta'\cH+2\eta\eta'']
+4a^2\left(\sigma_{00}+\frac{1}{2}\sigma_0^2\right)\nonumber\\
&&+\frac{a^2}{\eta}\left(-\eta'+\eta\cH\right)\sigma_0
-\frac{1}{2}\sum_{k=1}^3\sigma_{kk}-\sum_{k=1}^3\sigma_k^2.
\eea

\section{The generalized Friedmann equations, and their cosmological implications}\label{sect3}

We {\it postulate} now that the Einstein gravitational field equations can be formulated in the conformal Barthel-Kropina geometry in the form
\be\label{Ein}
\hat{\tilde{G}}_{IJ}=\frac{8\pi G}{c^4}\hat{\tilde{T}}_{IJ},
\ee
where $\hat{\tilde{T}}_{IJ}$ is the matter energy-momentum tensor in the conformal frame. Similarly to the previous investigations,  we assume that the thermodynamic properties of the cosmological matter in the conformal Barthel-Kropina geometry are characterized by the energy density $\hat{\tilde{\rho}} c^2$, and the thermodynamic pressure $\hat{\tilde{p}}$ only. Moreover, {\it we assume the existence of a frame comoving with matter}.  Hence, {\it we postulate} that the energy-momentum tensor of the matter takes in the conformal frame the form
 \be
\hat{\tilde{T}}_{I}^{J}=\begin{pmatrix}
\hat{\tilde{\rho}} c^2 & 0 & 0 & 0\\
0 & -\hat{\tilde{p}} & 0 & 0\\
0 & 0 & -\hat{\tilde{p}} & 0\\
0 & 0 & 0 & -\hat{\tilde{p}}
\end{pmatrix},
\ee
and
 \be
\hat{\tilde{T}}_{IJ}=e^{-2\sigma (x)}\begin{pmatrix}
\frac{e^{2\sigma (x)}}{a^2\eta ^2}\hat{\tilde{\rho}} c^2 & 0 & 0 & 0\\
0 & \frac{2e^{2\sigma (x)}}{\eta ^2}\hat{\tilde{p}} & 0 & 0\\
0 & 0 & \frac{2e^{2\sigma (x)}}{\eta ^2}\hat{\tilde{p}} & 0\\
0 & 0 & 0 & \frac{2e^{\sigma (x)}}{\eta ^2}\hat{\tilde{p}}
\end{pmatrix},
\ee
respectively.

Due to the homogeneity and isotropy of the cosmological space-time, all physical and geometrical quantities can depend only on the time coordinate $x^0$. As for the conformal factor, we assume first that it has the form,
\be
\sigma (x)=\phi \left(x^0\right)+\gamma _1x+\gamma _2y+\gamma _3z,
\ee
where $\gamma_i$, $i=1,2,3$ are arbitrary constants. For this form of the conformal factor we have $\sigma _0=\phi '\left(x^0\right)$, $\sigma _{00}=\phi '' \left(x^0\right)$, $\sigma _i=\gamma _i$,  and $\sigma _{ij}\equiv 0$, $i,j=1,2,3$, respectively. With this choice, Eq.~(\ref{41}) gives $\hat{\tilde{G}}_{ij}
=-\gamma_i\gamma_j=0$, which implies $\gamma _i=0$, $i=1,2,3$. Hence, we will chose the conformal factor as $\sigma (x)=\phi \left(x^0\right)$, and thus {\it we will restrict our analysis to the time dependent only conformal transformations of the Kropina metric}.

\subsection{The generalized Friedmann equations}

Then, the generalized Friedmann equations, describing the cosmological evolution in the conformal Barthel-Kropina geometry take the form
\be\label{B1}
\frac{3(\eta')^2}{\eta^2}=\frac{8\pi G}{c^2}\frac{1}{a^2\eta ^2}\hat{\tilde{\rho}}+3\left(\phi '\right)^2-6\frac{\eta '}{\eta }\phi ',
\ee
and
\bea\label{B2}
&&\frac{2}{\eta^2}[-3(\eta')^2+2\eta\eta'\cH+2\eta\eta'']
=\frac{16\pi G}{c^4}\frac{1}{a^2\eta ^2}\hat{\tilde{p}}\nonumber\\
&&-4\left[\phi ''+\frac{1}{2}\left(\phi '\right)^2\right]
+\left(\frac{\eta'}{\eta}-\cH\right)\phi',
\eea
respectively. By eliminating the term $-3\left(\eta '\right)^2/\eta ^2$ between Eqs.~(\ref{B1}) and (\ref{B2}) we obtain the relation
\bea
2\frac{1}{a\eta}\frac{d}{dx^0}\left(a\eta '\right)&=&\frac{4\pi G}{c^4}\frac{1}{a^2\eta ^2}\left(\hat{\tilde{\rho}}c^2+\hat{\tilde{p}}\right)-\left(\phi ''-\left(\phi '\right)^2\right)\nonumber\\
&&-\frac{11}{4}\frac{\eta '}{\eta} \phi'
-\frac{1}{4}\phi ' \cH.
\eea

We consider now the limiting case of the system (\ref{B1})-(\ref{B2}), corresponding to $\eta \rightarrow 1/a$, $\left(A_I(x)\right)=(1,0,0,0)$, and $\beta =y^0$. Hence, the generalized Friedmann equations of the conformal Barthel-Kropina model take the form
\be
3\cH ^2=\frac{8\pi G}{c^4}\hat{\tilde{\rho}}c^2+3\left(\phi '\right)^2+6\cH \phi',
\ee
and
\be
2\cH '+3\cH ^2=-\frac{8\pi G}{c^4}\hat{\tilde{p}}+2\left[\phi ''+\frac{1}{2}\left(\phi '\right)^2\right]+\cH \phi',
\ee
respectively. For $\phi=0$ we fully recover the standard Friedmann equations of general relativity.

\subsubsection{The de Sitter solution}

In the case of the standard Barthel-Kropina cosmological model, with $\phi =0$,  there is no vacuum, de Sitter type solution, of the field equations, since for $\rho =p=0$ all field equations are satisfied by the simple case $\eta '=0$, $\eta ={\rm constant}$, a solution independent of the concrete form of $\cH$. The situation is different in the conformal Barthel-Kropina model. By assuming  $\hat{\tilde{\rho}}=\hat{\tilde{p}}=0$, Eq.~(\ref{B1}) can be reformulated as
\be
\left(\phi' -\frac{\eta '}{\eta}\right)^2=2\frac{\left(\eta '\right)^2}{\eta ^2},
\ee
giving
\be
\phi'=\left(1\pm \sqrt{2}\right)\frac{\eta '}{\eta}=\varsigma \frac{\eta'}{\eta},
\ee
where we have denoted $\varsigma =1\pm \sqrt{2}$. Then, for $\cH=\cH_0={\rm constant}$,  Eq.~(\ref{B2}) takes the form
\bea
&&4 \varsigma (\varsigma+1) \phi ''\left(x^0\right)+\left(2 \varsigma ^2+\varsigma-2\right) \phi'\left(x^0\right)^2\nonumber\\
&&-(\varsigma-4) \varsigma \cH_0 \phi '\left(x^0\right)=0,
\eea
having, for $\varsigma =1+\sqrt{2}$, the general solution given by
\bea
\phi \left( x^{0}\right) &=&c_{2}-\frac{8\left( 3-\sqrt{2}\right) }{5\left( 8-5%
\sqrt{2}\right) }\times \nonumber\\
&&\Bigg\{ \ln \left[ 1+5e^{\left( 4\sqrt{2}-5\right)
c_{1}\cH_{0}}e^{\left( \frac{5\sqrt{2}}{8}-1\right) \cH_{0}x^{0}}\right]
-5c_{1}\cH_{0}\Bigg\},\nonumber\\
\eea
where $c_1$ and $c_2$ are arbitrary constants of integration. In the limit of large times $\phi \left(x^0\right)$ tends to a constant. During the vacuum de Sitter phase of expansion, the function $\eta $ is given by $\eta \propto e^{\phi /\varsigma}$.

\subsection{Dark energy in the conformal Barthel-Kropina model}

We will investigate now the possibility of the dark energy description as a geometric effect in the Barthel-Kropina cosmological model. We have already seen that in the limit $\eta\rightarrow 1/a$, and $\phi =0$, the general relativistic model without a cosmological constant is recovered. We will assume now that the departures from general relativity can be described by a small variation of $\eta$, which depend on the conformal factor. Hence, tentatively we propose a cosmological model in which $\eta$ has the form
\be\label{eta}
\eta =\frac{e^{\gamma \phi}}{a},
\ee
where $\gamma $ is a constant. Thus, we immediately obtain
\be
\frac{\eta '}{\eta}=\gamma \phi '-\cH, \frac{\eta ''}{\eta}=\gamma \phi ''-\cH '+\left(\gamma \phi '-\cH\right)^2.
\ee

Hence, the generalized Friedmann equations (\ref{B1}) and (\ref{B2}) of  the conformal Barthel-Kropina model become
\bea\label{58}
3\cH ^2&=&\frac{8\pi G}{c^2}e^{-2\gamma \phi}\hat{\tilde{\rho}}+3\left(1-2\gamma-\gamma ^2\right)\left(\phi '\right)^2+6(1+\gamma)\phi ' \cH \nonumber\\
&=&\frac{8\pi G}{c^2}e^{-2\gamma \phi}\hat{\tilde{\rho}}+\hat{\tilde{\rho}}_\phi,
\eea
and
\bea\label{59}
2\cH '+3\cH ^2&=&-\frac{8\pi G}{c^4}e^{-2\gamma \phi}\hat{\tilde{p}}+2(1+\gamma)\phi '' \nonumber\\
&&-\left(1-\frac{1}{2}\gamma -\gamma ^2\right)\left(\phi '\right)^2
+(1+4\gamma)\phi '\cH\nonumber\\
&&=-\frac{8\pi G}{c^4}e^{-2\gamma \phi}\hat{\tilde{p}}-\hat{\tilde{p}}_\phi,
\eea
respectively, where we have denoted
\be
\hat{\tilde{\rho}}_\phi=3\left(1-2\gamma-\gamma ^2\right)\left(\phi '\right)^2+6(1+\gamma)\phi ' \cH,
\ee
and
\be
\hat{\tilde{p}}_\phi=-2(1+\gamma)\phi '' +\left(1-\frac{1}{2}\gamma -\gamma ^2\right)\left(\phi '\right)^2-(1+4\gamma)\phi '\cH,
\ee
respectively.

From Eqs.~(\ref{58}) and (\ref{59}) we obtain immediately
\bea\label{62}
\hspace{-0.5cm}\cH '&=&-\frac{4\pi G}{c^4}e^{-2\gamma \phi}\left(\hat{\tilde{\rho}}c^2+\hat{\tilde{p}}\right)+(1+\gamma)\phi ''\nonumber\\
\hspace{-0.5cm}&&+\frac{1}{2}\left(4\gamma ^2+\frac{13}{2}\gamma-4\right)\left(\phi '\right)^2-\frac{1}{2}(2\gamma +5)\phi '\cH.
\eea

Eqs.~(\ref{58}) and (\ref{59}) give the total conservation equation for matter and the conformally induced scalar field as
\bea\label{63}
&&\frac{8\pi G}{c^4}\left[\hat{\tilde{\rho}}'+3\cH\left(\hat{\tilde{\rho}}+\frac{\hat{\tilde{p}}}{c^2}\right)-2\gamma \phi ' \hat{\tilde{\rho}} \right]e^{-2\gamma \phi}\nonumber\\
&&+\hat{\tilde{\rho}}'_\phi+3\cH \left(\hat{\tilde{\rho}}_\phi+\hat{\tilde{p}}_\phi\right)=0.
\eea

We split Eq.~(\ref{63}) into two balance equations, for matter, and the conformal scalar field, respectively, given by
\be\label{64}
\hat{\tilde{\rho}}'+3\cH\left(\hat{\tilde{\rho}}+\frac{\hat{\tilde{p}}}{c^2}\right)-2\gamma \phi ' \hat{\tilde{\rho}} =0,
\ee
and
\be\label{65}
\hat{\tilde{\rho}}'_\phi+3\cH \left(\hat{\tilde{\rho}}_\phi+\hat{\tilde{p}}_\phi\right)=0,
\ee
respectively. Eq.~(\ref{65}) gives the dynamical evolution of the conformal scalar field as
\bea\label{66}
&&6\left(1-2\gamma -\gamma ^2\right)\phi ''+3\left(4-\frac{13}{2}\gamma -4\gamma ^2\right)\cH\phi' \nonumber\\
&& +6(1+\gamma)\cH '+3(2\gamma +5)\cH ^2=0.
\eea

Hence, we have obtained the basic equations describing the cosmological dynamics in the conformal Barthel-Kropina model as given by Eqs.~(\ref{62}), (\ref{64}) and (\ref{66}), respectively.

\section{Cosmological models in the conformal Barthel-Kropina theory}\label{sect4}

In the present Section we will investigate the cosmological viability of the conformal Barthel-Kropina cosmological model. More exactly, we will consider the late time evolution of the model, and we will compare its predictions with a selected set of cosmological data for the Hubble function. We will assume that the matter content of the present day Universe can be well described by a pressureless fluid, with zero pressure, and hence in the following we will take $\hat{\tilde{p}}=0$.

\subsection{Cosmological evolution equations}

From Eqs.~(\ref{62}) and (\ref{66}) we obtain $\cH '$ and $\phi ''$ in the form
\bea
\cH '&=&\frac{1}{4}\left( \gamma ^{2}+2\gamma -1\right) \frac{8\pi G}{c^{2}}\hat{\tilde{\rho}}
e^{-2\gamma \phi }\nonumber\\
&&-\frac{1}{8}\left( 8\gamma ^{4}+29\gamma ^{3}+10\gamma
^{2}-29\gamma +8\right) \left( \phi ^{\prime }\right) ^{2}\nonumber\\
&&-\frac{1}{4}\left(
2\gamma ^{2}+7\gamma +5\right) \cH^{2}\nonumber\\
&&+\frac{3}{8}\left( 4\gamma ^{3}+13\gamma
^{2}+7\gamma -6\right) \cH\phi ^{\prime },\
\eea
and
\bea
\phi ^{\prime \prime }&=&\frac{1}{4}(\gamma +1)\frac{8\pi G}{c^{2}}\hat{\tilde{\rho}}
e^{-2\gamma \phi }\nonumber\\
&&-\frac{1}{8}\left( 8\gamma ^{3}+21\gamma ^{2}+5\gamma
-8\right) \left( \phi ^{\prime }\right) ^{2}\nonumber\\
&&-\frac{1}{4}(2\gamma +5)\cH^{2}+%
\frac{1}{8}\left( 12\gamma ^{2}+27\gamma +2\right) \cH\phi ^{\prime },
\eea
respectively.

\subsubsection{The dimensionless form of the generalized Friedmann equations}

We will replace now in the evolution equations the coordinate $x^0=ct$ with the time $t$, and the Hubble function $\cH$ by the time dependent Hubble function $H=\dot{a}/a$, where a dot denotes the derivative with respect to the time $t$, so that $\cH=H/c$. We will also introduce a set of dimensionless variables $\left(h,\tau,r_m\right)$, defined according to
\be
H=H_0h, \tau =H_0t, \hat{\tilde{\rho}}=\frac{3H_0^2}{8\pi G}r_m,
\ee
where $H_0$ is the present day value of the Hubble function. Hence, we obtain the full set of the evolution equations of the conformal Barthel-Kropina cosmological model as
\be\label{70}
\frac{dr_m}{d\tau}+3hr_m=2\gamma \frac{d\phi}{d\tau}r_m,
\ee
\begin{eqnarray}
\frac{dh}{d\tau } &=&\frac{3}{4}\left( \gamma ^{2}+2\gamma -1\right)
r_{m}e^{-2\gamma \phi }  \nonumber \\
&&-\frac{1}{8}\left( 8\gamma ^{4}+29\gamma +10\gamma ^{2}-29\gamma
+8\right) \left( \frac{d\phi }{d\tau }\right) ^{2}  \nonumber \\
&&-\frac{1}{4}\left( 2\gamma ^{2}+7\gamma +5\right) h^{2}\nonumber\\
&&+\frac{3}{8}\left(
4\gamma ^{3}+13\gamma ^{2}+7\gamma -6\right) h\frac{d\phi }{d\tau },
\end{eqnarray}%
and
\begin{eqnarray}
\hspace{-0.5cm}\frac{d^{2}\phi }{d\tau ^{2}} &=&\frac{3}{4}(\gamma +1)r_{m}e^{-2\gamma \phi
}  \nonumber \\
\hspace{-0.5cm}&&-\frac{1}{8}\left( 8\gamma ^{3}+21\gamma ^{2}+5\gamma -8\right) \left(
\frac{d\phi }{d\tau }\right) ^{2}  \nonumber \\
\hspace{-0.5cm}&&-\frac{1}{4}(2\gamma +5)h^{2}+\frac{1}{8}\left( 12\gamma ^{2}+27\gamma
+2\right) h\frac{d\phi }{d\tau },
\end{eqnarray}%
respectively.

Eq.~(\ref{70}) for the matter density can be immediately integrated to give
\be
r_m(\tau)=r_{m0}\frac{e^{2\gamma \phi}}{a^3}=r_{m0}\frac{\eta ^2}{a},
\ee
where $r_{m0}$ is an arbitrary constant of integration.

\subsubsection{The redshift representation}

In order to facilitate the comparison with the observational data we introduce the redshift variable $z$ defined as $1+z=1/a$, giving $d/d\tau =-(1+z)h(z)d/dz$. Hence, we can reformulate the cosmological evolution equations in the redshift space as
\be\label{F1}
-(1+z)h\frac{d\phi}{dz}=u,
\ee
\be\label{F2}
(1+z)\frac{dr_m}{dz}-3r_m=2\gamma (1+z)\frac{d\phi}{dz}r_m,
\ee
\begin{eqnarray}\label{F3}
-(1+z)h\frac{dh}{dz} &=&\frac{3}{4}\left( \gamma ^{2}+2\gamma -1\right)
r_{m}e^{-2\gamma \phi }  \nonumber \\
&&-\frac{1}{8}\left( 8\gamma ^{4}+29\gamma ^{3}+10\gamma ^{2}-29\gamma
+8\right) u^{2}  \nonumber \\
&&-\frac{1}{4}\left( 2\gamma ^{2}+7\gamma +5\right) h^{2}\nonumber\\
&&+\frac{3}{8}\left(
4\gamma ^{3}+13\gamma ^{2}+7\gamma -6\right) hu,
\end{eqnarray}
and
\begin{eqnarray}\label{F4}
-(1+z)h\frac{du}{dz} &=&\frac{3}{4}(\gamma +1)r_{m}e^{-2\gamma \phi }
\nonumber \\
&&-\frac{1}{8}\left( 8\gamma ^{3}+21\gamma ^{2}+5\gamma -8\right) u ^{2}  \nonumber \\
&&-\frac{1}{4}(2\gamma +5)h^{2}+\frac{1}{8}\left( 12\gamma ^{2}+27\gamma
+2\right) hu, \nonumber\\
\end{eqnarray}
respectively. The system of equations (\ref{F1})-(\ref{F4}) must be integrated with the initial conditions $h(0)=1$, $\phi (0)=\phi_0$, $u(0)=u_0$, and $r_m(0)=r_{m0}$, respectively. We also introduce the deceleration parameter $q$, an indicator of the decelerating/accelerating nature of the cosmological expansion, defined as
\be
q=\frac{d}{d\tau}\frac{1}{h}-1=-(1+z)h\frac{d}{dz}\frac{1}{h}-1=(1+z)\frac{1}{h}\frac{dh}{dz}-1.
\ee
In order to test the viability of the conformal Barthel-Kropina cosmological model we will compare its predictions with the results of the $\Lambda$CDM standard model. In the $\Lambda$CDM model, the redshift evolution of the Hubble function $H$ is given by
\be
H(z)=H_0\sqrt{\Omega _m^{(cr)}(1+z)^3+\Omega _\Lambda},
\ee
where $\Omega _m^{(cr)}=\Omega _b^{(cr)}+\Omega _{DM}^{(cr)}$, where $\Omega _b^{(cr)}$ and $\Omega _{DM}^{(cr)}$ denote the critical densities of the baryonic and dark matter, defined generally as $\Omega _i^{(cr)}=\rho _i/\rho_{cr}$, $i=b, DM$, where $\rho_{cr}=3H_0^2/8\pi G$. The density parameter of the dark energy (a cosmological constant) is defined as $\Omega _\Lambda =\Lambda/\rho_{cr}$. The deceleration parameter is given by
\be
q(z)=\frac{3(1+z)^3\Omega _m^{(cr)}}{2\left[\Omega _\Lambda+(1+z)^3\Omega _m^{(cr)}\right]}-1.
\ee
For the matter and dark energy density parameters we adopt the numerical values
$\Omega_{DM}^{(cr)} = 0.2589$, $\Omega b^{(cr)} = 0.0486$, and $\Omega _\Lambda = 0.6911$, respectively \cite{Planck1, Planck2}. This gives for the total matter density parameter
$\Omega _m^{(cr)}=\Omega _b^{(cr)}+\Omega _{DM}^{(cr)}$ the value $\Omega _m^{(cr)} = 0.3089$. The present
day value of the deceleration parameter is
$q(0) =-0.5381$, a value that indicates that the Universe is presently
in an accelerating epoch.

\subsection{The case $\gamma =1$}

We will begin the analysis of the cosmological implications of the conformal Barthel-Kropina model by considering the simple case corresponding to $\gamma =1$. For this particular choice of the model parameter the field equations take the form
\be\label{F11}
-(1+z)h\frac{d\phi}{dz}=u,
\ee
\be\label{F21}
(1+z)\frac{dr_m}{dz}-3r_m=2 (1+z)\frac{d\phi}{dz}r_m,
\ee
\bea\label{F31}
-(1+z)h\frac{dh}{dz}=r_me^{-2\phi}-\frac{7}{2}h^2+\frac{27}{4}hu-\frac{13}{4}u^2,
\eea
\bea\label{F41}
-(1+z)h\frac{du}{dz}=r_me^{-2\phi}-\frac{7}{4}h^2+\frac{41}{8}hu-\frac{13}{4}u^2.
\eea

\begin{figure*}[htbp]
\includegraphics[scale=0.6]{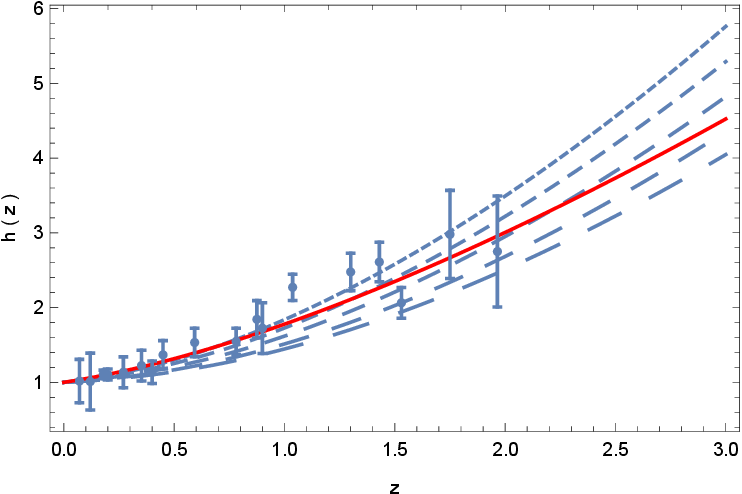}
\includegraphics[scale=0.6]{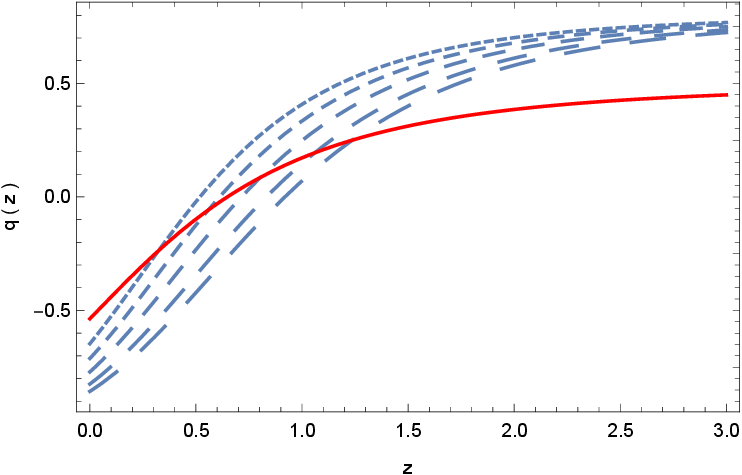}
\caption{Variation of the dimensionless Hubble function $h$ (left panel) and of the deceleration parameter (right) panel in the conformal Barthel-Kropina cosmological model for $\gamma =1$, and for $u(0)=0.69$ (dotted curve), $u(0)=0.72$ (short dashed curve), $u(0)=0.75$ (dashed curve), $u(0)=0.78$ (long-dashed curve), and $u(0)=0.80$ (ultra-long dashed curve). The initial conditions used to integrate the cosmological evolution equations are $\phi (0)=0.16$, $r_m(0)=0.05$, and $h(0)=1$, respectively. The observational data are represented with their error bars, while the red curve depicts the predictions of the $\Lambda$CDM model.}\label{fig1}
\end{figure*}

The system of equations (\ref{F11})-(\ref{F41}) must be integrated with the initial conditions $\phi (0)=\phi_0$, $u(0)=u_0$, $r_m(0)=r_{m0}$, and $h(0)=1$, respectively. The variations with respect to the redshift of the dimensionless Hubble function and of the deceleration parameter are represented in Fig.~\ref{fig1}.

\begin{figure*}[htbp]
\includegraphics[scale=0.6]{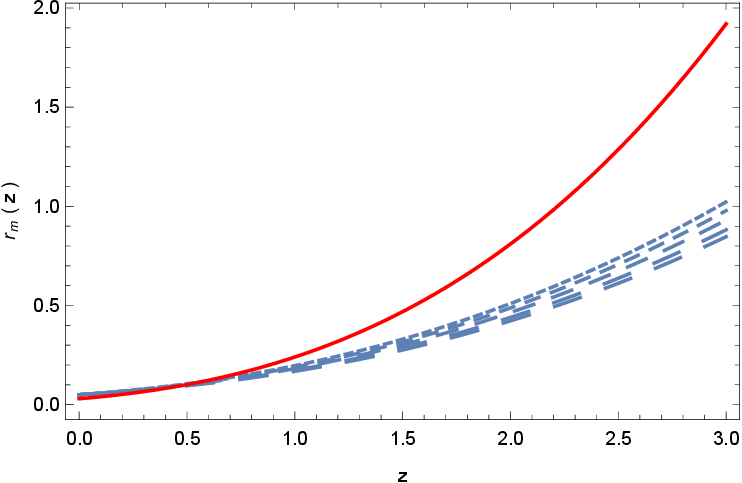}
\includegraphics[scale=0.6]{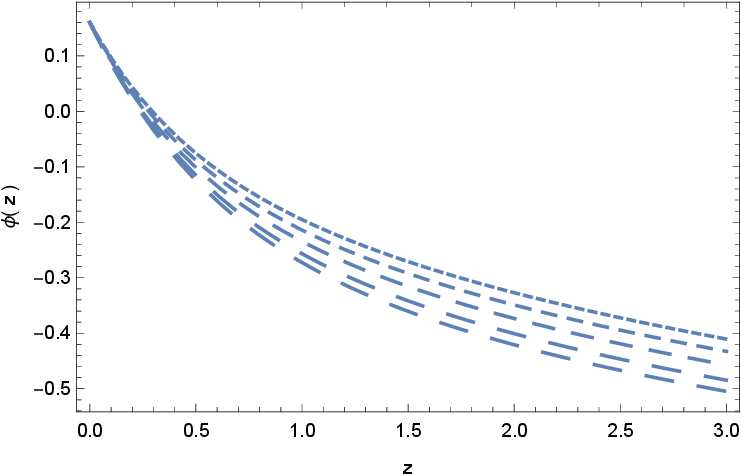}
\caption{Variation of the dimensionless matter density  $r$ (left panel) and of the conformal factor $\phi$ (right) panel in the conformal Barthel-Kropina cosmological model for $\gamma =1$, and for $u(0)=0.69$ (dotted curve), $u(0)=0.72$ (short dashed curve), $u(0)=0.75$ (dashed curve), $u(0)=0.78$ (long-dashed curve), and $u(0)=0.80$ (ultra-long dashed curve). The initial conditions used to integrate the cosmological evolution equations are $\phi (0)=0.16$, $r_m(0)=0.05$, and $h(0)=1$, respectively. The red curve shows the predictions of the $\Lambda$CDM model.}\label{fig2}
\end{figure*}

The variations of the matter energy density and of the conformal factor $\phi$ are represented in Fig.~\ref{fig2}.

As one can see from Fig.~\ref{fig1}, the conformal Barthel-Kropina model gives a good description of the observational data for the Hubble function, and it almost perfectly overlaps with the prediction of the $\Lambda$CDM model up to a redshift of $z=2.5$. For higher redshifts some important differences between models may appear. On the other hand, significant differences do exist in the behavior of the deceleration parameter, the conformal Barthel-Kropina model predicting higher deceleration values at higher redshifts, and lower values at small redshifts. The baryonic matter content, presented in Fig.~\ref{fig2},  appears to be much higher in the $\Lambda$CDM model. In fact, in the conformal Barthel-Kropina model less baryonic matter is predicted to exist at higher redhifts. The conformal factor $\phi(z)$ evolves from positive values at small redshifts to larger negative values, increasing rapidly with $z$.

\subsection{The case $\gamma =-1$}

\begin{figure*}[htbp]
\includegraphics[scale=0.6]{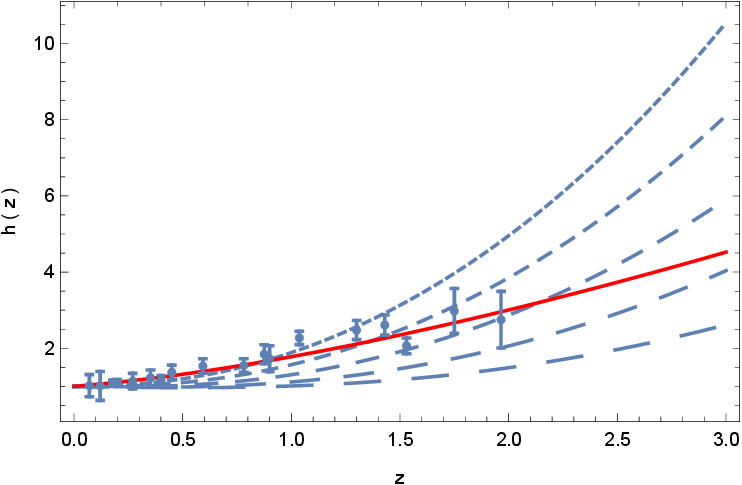}
\includegraphics[scale=0.6]{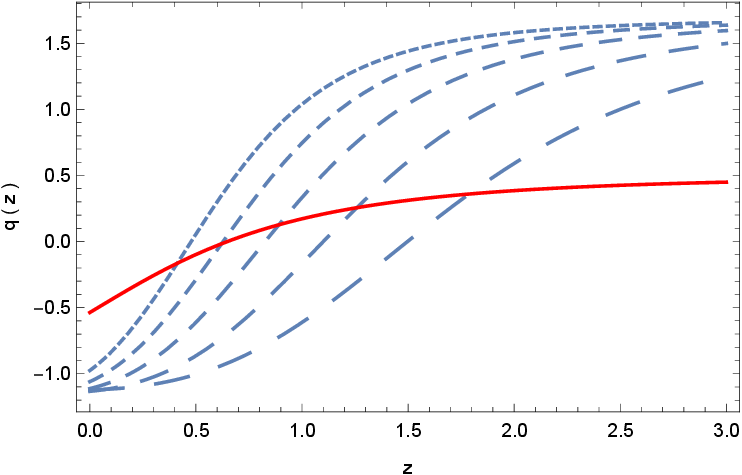}
\caption{Variation of the dimensionless Hubble function $h$ (left panel) and of the deceleration parameter (right) panel in the conformal Barthel-Kropina cosmological model for $\gamma =-1$, and for $u(0)=-0.01$ (dotted curve), $u(0)=-0.08$ (short dashed curve), $u(0)=-0.15$ (dashed curve), $u(0)=-0.22$ (long-dashed curve), and $u(0)=-0.29$ (ultra-long dashed curve). The initial conditions used to integrate the cosmological evolution equations are $\phi (0)=0.025$, $r_m(0)=0.05$, and $h(0)=1$, respectively. The observational data are represented with their error bars, while the red curve depicts the predictions of the $\Lambda$CDM model.}\label{fig3}
\end{figure*}

For $\gamma =-1$, the variations of the Hubble function and of the deceleration parameter are presented in Fig.~\ref{fig3}.

For negative values of $\gamma$, the conformal Barthel-Kropina model generally does not give a particularly good description of the observational data, or reproduces closely the $\Lambda$CDM model. As one can see from the left panel of Fig.~\ref{fig3}, important differences may appear between observations at both small and high redshifts. The differences are even more important in the case of the deceleration parameter, with the conformal Barthel-Kropina cosmology always ending at the present time in a de Sitter type accelrating, or super-accelerating phase, with $q$ equal, or smaller than -1. At high redshifts, the deceleration parameter takes much higher values than those predicted by the $\Lambda$CDM model.

\begin{figure*}[htbp]
\includegraphics[scale=0.6]{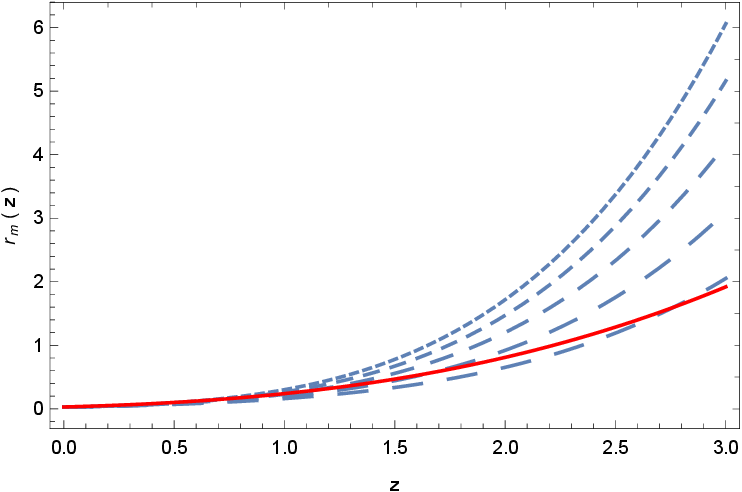}
\includegraphics[scale=0.6]{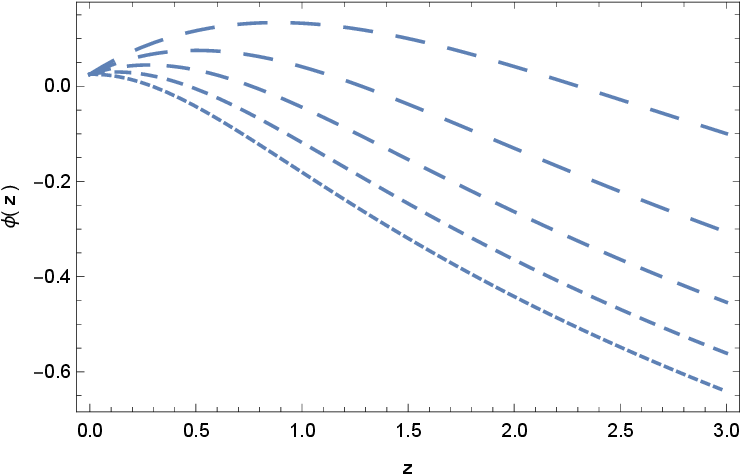}
\caption{Variation of the dimensionless matter density $r$ (left panel) and of the conformal factor $\phi$  (right) panel in the conformal Barthel-Kropina cosmological model for $\gamma =-1$, and for $u(0)=-0.01$ (dotted curve), $u(0)=-0.08$ (short dashed curve), $u(0)=-0.15$ (dashed curve), $u(0)=-0.22$ (long-dashed curve), and $u(0)=-0.29$ (ultra-long dashed curve). The initial conditions used to integrate the cosmological evolution equations are $\phi (0)=0.025$, $r_m(0)=0.05$, and $h(0)=1$, respectively. The red curve shows the predictions of the $\Lambda$CDM model.}\label{fig4}
\end{figure*}

The variations of the matter density and of the conformal factor $\phi$ are represented in Fig.~\ref{fig4}.  While at low redshifts, up to the order of $z\approx 1.5$, the predicted matter density of the conformal Barthel-Kropina model has similar values as in the $\Lambda$CDM model, at higher redshifts a much higher matter density is predicted as compared to standard cosmology. This different behavior can be traced back to the matter density evolution equation (\ref{F2}), which indicates the possibility of the nonconservation of the baryonic content of the Universe. The conformal factor $\phi$, shown in the right panel of Fig.~\ref{fig4}), increases initially with the redshift, reaching a maximum value at $z\in (0.4,1)$. Then, the conformal factor decreases, and acquires negative values at higher redshifts.

\subsection{The case $\gamma =4/3$}

\begin{figure*}[htbp]
\includegraphics[scale=0.6]{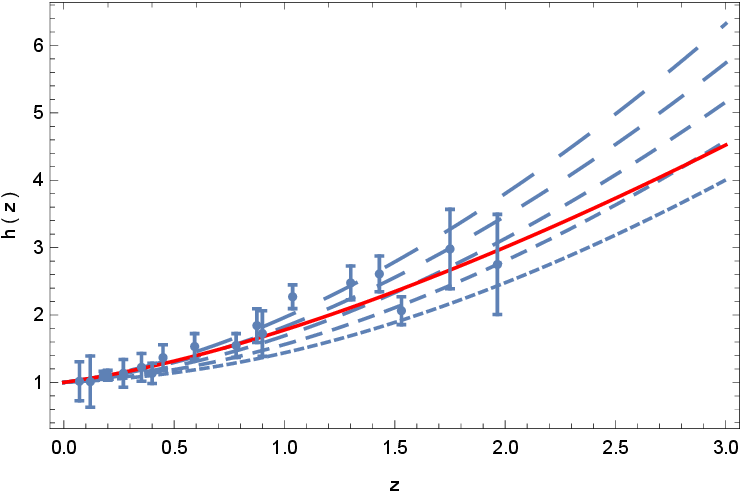}
\includegraphics[scale=0.6]{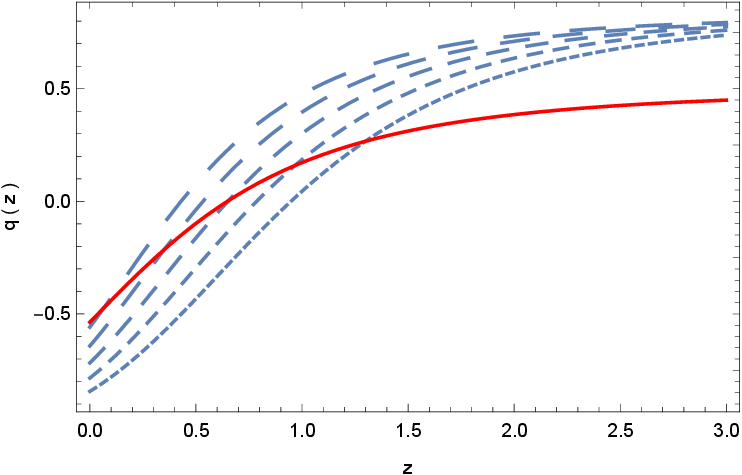}
\caption{Variation of the dimensionless Hubble function $h$ (left panel) and of the deceleration parameter (right) panel in the conformal Barthel-Kropina cosmological model for $\gamma =4/3$, and for $u(0)=0.80$ (dotted curve), $u(0)=0.82$ (short dashed curve), $u(0)=0.84$ (dashed curve), $u(0)=0.86$ (long-dashed curve), and $u(0)=0.88$ (ultra-long dashed curve). The initial conditions used to integrate the cosmological evolution equations are $\phi (0)=0.56$, $r_m(0)=0.05$, and $h(0)=1$, respectively. The observational data are represented with their error bars, while the red curve represents the predictions of the $\Lambda$CDM model.}\label{fig5}
\end{figure*}

Finally, we consider the cosmological evolution of the homogeneous and isotropic conformal Barthel-Kropina cosmological model for $\gamma =4/3$.  The variations of the Hubble function and of the deceleration parameter are presented in Fig.~\ref{fig5}, together with the observational data, and the predictions of the $\Lambda$CDM cosmological model. The conformal Barthel-Kropina model can give a satisfactory description of  the observational data, and for a certain range of the numerical values of the model parameters (the initial conditions for the evolution equation of the conformal factor), it can reproduces almost exactly the predictions of the $\Lambda$CDM paradigm. However, the important differences in the predictions on the behavior of the deceleration parameter still persist, indicating lower values at small redshifts, and higher values at high redshifts. A de Sitter type evolution, with $q(0)=-1$ is also obtained for this value of $\gamma$. Parameter values that give at low redshifts the same evolution of $q$ as in $\Lambda$CDM still predict much higher values at higher redshifts.

\begin{figure*}[htbp]
\includegraphics[scale=0.6]{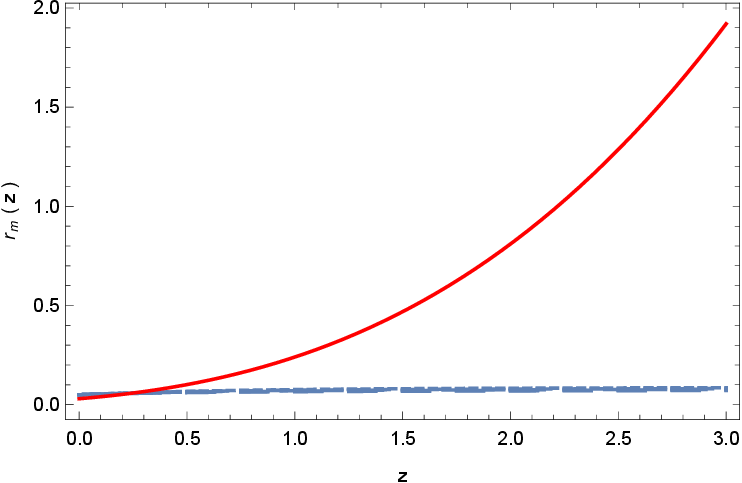}
\includegraphics[scale=0.6]{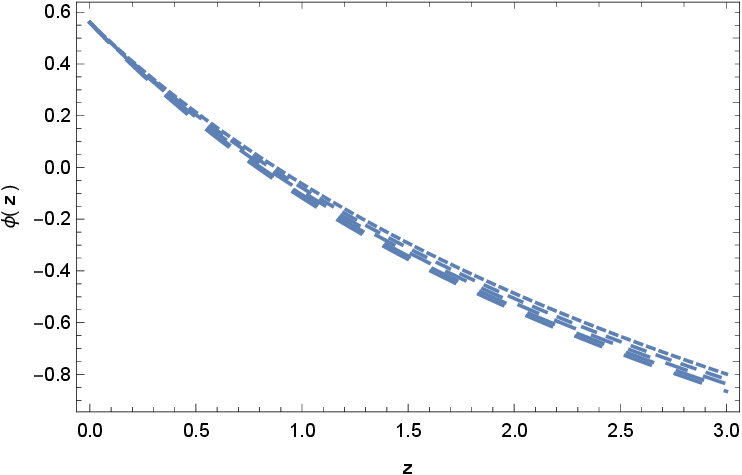}
\caption{Variation of the dimensionless matter density  $r$ (left panel) and of the conformal factor $\phi$ (right) panel in the conformal Barthel-Kropina cosmological model for $\gamma =4/3$, and and for $u(0)=0.80$ (dotted curve), $u(0)=0.82$ (short dashed curve), $u(0)=0.84$ (dashed curve), $u(0)=0.86$ (long-dashed curve), and $u(0)=0.88$ (ultra-long dashed curve). The initial conditions used to integrate the cosmological evolution equations are $\phi (0)=0.56$, $r_m(0)=0.05$, and $h(0)=1$, respectively. The red curve represents the predictions of the $\Lambda$CDM model.}\label{fig6}
\end{figure*}

The redshift evolutions  of the baryonic matter energy density and of the conformal factor $\phi$ are represented for the $\gamma =4/3$ conformal Barthel-Kropina model in Fig.~\ref{fig6}.

At redshifts $z>0.5$, predictions of the baryonic matter content of the $\Lambda$CDM model largely exceed the predictions of the conformal Barthel-Kropina model, whose matter content is relatively constant, and equal to the present day value. This aspect may be again explained by the conformal dependence of the matter density on the conformal factor, as cen be seen in Eq.~(\ref{F2}). The conformal factor $\phi$ monotonically decreases from its initial, positive value, to negative values reached at higher redshifts. The variations of the basic model quantities $\left(r_m,h,\phi\right)$ do not depend significantly on the initial values of the conformal factor $\phi (0)$, but shows a strong dependence on the initial values of its derivative $\left.\left(d\phi(z)/dz\right)\right|_{z=0}$, which gives the rate of the variation of the conformal factor with the cosmological redshift.

\section{Discussions and final remarks}\label{sect5}

In the present paper we have considered an extension of the Barthel-Kropina dark energy model, as introduced in \cite{Fc27} and (\cite{Fc28}, by introducing the mathematical and physical perspective of the conformal transformations. Conformal transformations are the basis of the Weyl geometry \cite{We1,We2}, interestingly proposed in the same year, 1918, as Finsler geometry \cite{F1}. However, the relations between these two fundamentals geometries have been explored very little, if at all, in a physical context. However, there are some mathematical studies trying to fill de gap between these geometries, and investigating the effects of the conformal transformations on various Finsler type geometries \cite{Conf0,Conf1,Conf2,Conf3,Conf4,Conf5,Conf6,Conf7}.

The starting point of our investigation is Eq.~(\ref{Cftrans}), in which we have assumed that the fundamental function $F(x,y)$ of a Finsler geometry is conformally transformed into a new function $\tilde{F}(x,y)$. In the case of an $(\alpha, \beta)$ metric, a conformal transformation acts on both the Riemannian metric $\alpha$, and on the one-form $\beta$, as indicated by Eq.~(\ref{Cftrans1}). We can interpret, by analogy with standard Riemannian general relativity,  such a conformal transformation as a transformation from a Finsler-type Einstein frame to a Finsler-type Jordan frame. Mathematically, such a transformation is always possible, and leads to some classes of Finsler geometries with interesting properties. From a mathematical point of view one can define a conformal change in the Finsler geometry as follows. Let $F^n=\left(M^n.L\right)$ and $\tilde{F}^n=\left(M^n.\tilde{L}\right)$ two Finsler spaces defined on the same underlying base manifold $M^n$, where by $L$ we have denoted the Finsler metric function. If the angle between any two tangent vectors in $F^n$ is equal to the angle in $\tilde{F}^n$, then $F^n$ is called conformal to $\tilde{F}^
n$, and the transformation $L\rightarrow \tilde{L}$ is called a conformal transformation of the metric. In other words, if there exists a scalar function $\sigma (x)$ such that $\tilde{L}=e^{\sigma (x)}L$, then the transformation is called a conformal transformation \cite{Conf4,Conf5}.

For the case of an $(\alpha, \beta)$ metric, $\tilde{L}=e^{\sigma (x)}L$ is equivalent to $\tilde{L}=\left(e^{\sigma (x)}\alpha, e^{\sigma (x)}\beta\right)$, or, equivalently, $\tilde{g}_{IJ}=e^{2\sigma (x)}g_{IJ}$, and $\tilde{A}_I=e^{\sigma (x)}A_I$. The conformal transformation of the connection in Riemann geometry is given by Eq.~(\ref{Lema1}), and it adds to the standard Levi-Civita connection three new terms determined by the derivatives of the conformal factor $\sigma$.

The original Barthel-Kropina cosmological model was built upon three fundamental mathematical assumptions. Firstly, we assume that the Finsler metric function is a Kropin type $(\alpha, \beta)$ metric. Secondly, the osculating approach to the Finsler-Kropinsa metric was adopted, in which $g)(x,y)$ becomes a Riemannian metric $g(x,Y(x))$. Thirdly, the first two assumptions lead to the result that the connection of this Riemann metric is nothing but the corresponding Levi-Civita connection, called, in Finsler geometry, the Barthel connection. With the help of these three assumptions one can construct systematically the corresponding Einstein gravitational field equations, and, in a cosmological setting, the generalized Friedmann equations that lead to a consistent cosmological model, which can be successfully tested against the cosmological observations. In the present study we have considered the conformal transformation of this model from its Einstein frame to the conformally related Jordan frame, which results in the introduction of a new scalar degree of freedom.

The corresponding cosmological model, constructed by assuming the homogeneous and isotropic Friedmann-Lemaitre-Robertson-Walker metric, leads to a set of generalized Friedmann equations, from which a dark energy effective geometric model can be constructed. There are two essential assumptions in the formulation of the model, the first being the adoption of the specific form (\ref{eta}) for the coefficient of the one-form $\beta$, which is considered to be related to the time-dependent conformal factor $\phi$ by an exponential relation, also involving the scale factor. As the second assumption in formulating the dark energy model we have considered the splitting of the global energy balance equations into two independent equations, with the equation of conservation of the conformal scalar field formulated in a form similar to that of ordinary scalar fields in cosmology. However, it is important to point out that in the present approach the scalar field is of purely geometric origin. Once these two assumptions are adopted, one can obtain a cosmological model, representing a generalization of the standard $\Lambda$CDM model, in which the cosmological evolution is described by the set of the Friedmann equations, plus an evolution equation for the scalar field. The model has one free parameter $\gamma$, and must be studied numerically once the initial conditions for the scalar field and its derivative at the time origin are given. Once these parameters are fixed, a large number of cosmological models can be obtained.

In our investigation we have restricted our analysis to three classes of models, determined by three distinct numerical values of the parameter $\gamma$. The initial conditions for the scalar field were slightly varied. To simplify the comparison with the observational data the redshift representation of the cosmological evolution equations was used. Moreover, we have compared the predictions of the conformal Barthel-Kropina model with the similar predictions of the $\Lambda$CDM model, and with a small set of observational cosmological data. The parameter $\gamma$ was given rather different values, both positive and negative, and integer and fractional. respectively. The comparison with the observational data was performed by using a trial and error method, and no formal mathematical fitting procedures were used. As a first conclusion of our study we can infer that positive values of $\gamma$ give a better description of the observational data for the Hubble function, and of $\Lambda$CDM cosmology, as compared to the negative $\gamma$ values. However, even negative values of $\gamma$ are not completely ruled out by the observations. On the other hand, important differences between the predictions of various cosmological parameters in the conformal Barthel-Kropina model and of the $\Lambda$CDM model do appear. These differences are especially significant for the case of the deceleration parameter, and of the matter density. Even that the conformal Barhel-Kropina model describes well the transition to an accelerating expansion, and can even reproduce the observational present day value of the deceleration parameter, significant differences appear especially at higher redshifts. where the conformal Barthel-Kropina predicts a much higher rate of deceleration as compared to the $\Lambda$CDM model. The existence of direct observational data on the deceleration parameter at higher redshifts would allow a precise discrimination between the cosmological models predicted by the conformal Barthel-Kropina model, and other standard of modified models. Other important differences do appear at the level of the ordinary matter density. For $\gamma =1$ and $\gamma =-1$, the conformal Barthel-Kropina model predicts higher matter densities, especially at redshifts $z>2$. On the other hand, for $\gamma =4/3$, the matter content of the Universe is much higher in the $\Lambda$CDM model than in the conformal Barthel-Kropina model, which predicts an almost constant matter density, showing a very small variation with the redshift.

To conclude, in the present study we have investigated an interesting feature of a specific Finsler type cosmology, related to the impact of a conformal transformation on the Kropina metric. The cosmological implications of the model have been considered in detail, and we have shown that this type of models may represent an attractive alternative to standard cosmologies based on Riemann geometries. We have developed the basic mathematical and physical tools that would allow the in depth comparison of the predictions of the conformal Barthel-Kropina model with the observational data, and with standard cosmological approaches. Hopefully the obtained results would lead to a better understanding the physical applications of Finsler geometry, and for its relevance for the description of the large scale cosmic phenomena, and processes.

\section*{Acknowledgments}

The work of TH is supported by a grant of the Romanian Ministry of Education
and Research, CNCS-UEFISCDI, project number PN-III-P4-ID-PCE-2020-2255
(PNCDI III).

\end{document}